\documentclass[prb]{revtex4}

\usepackage{graphicx}
\usepackage{dcolumn}
\usepackage{bm}

\begin{document}


\title{Phase control of La${}_{2}$CuO${}_{4}$ in thin-film synthesis}

\author{A. Tsukada}
\email{tsukada@with.brl.ntt.co.jp}
\affiliation{Department of Physics, Science University of Tokyo,
1-3, Kagurazaka, Shinjuku-ku, Tokyo 162-8601, Japan}

\author{T. Greibe}
\altaffiliation{On leave from Technical University of Denmark.}
\affiliation{NTT Basic Research Laboratories, NTT Corporation,\\
3-1, Morinosato-Wakamiya, Atsugi-shi, Kanagawa 243-0198, Japan}

\author{M. Naito}
\affiliation{NTT Basic Research Laboratories, NTT Corporation,\\
3-1, Morinosato-Wakamiya, Atsugi-shi, Kanagawa 243-0198, Japan}
\affiliation{Department of Physics, Science University of Tokyo,
1-3, Kagurazaka, Shinjuku-ku, Tokyo 162-8601, Japan}

\date{\today}

\begin{abstract}
The lanthanum copper oxide, La${}_{2}$CuO${}_{4}$, which is an end member of the
prototype high-$T$${}_{\rm c}$ superconductors (La,Sr)${}_{2}$CuO${}_{4}$ and
(La,Ba)${}_{2}$CuO${}_{4}$, crystallizes in the ``K${}_{2}$NiF${}_{4}$'' structure
in high-temperature bulk synthesis.  The crystal chemistry, however, predicts that
La${}_{2}$CuO${}_{4}$ is at the borderline of the K${}_{2}$NiF${}_{4}$
stability and that it can crystallize in the Nd${}_{2}$CuO${}_{4}$
structure at low synthesis temperatures.  In this article we demonstrate
that low-temperature thin-film synthesis actually crystallizes
La${}_{2}$CuO${}_{4}$ in the Nd${}_{2}$CuO${}_{4}$ structure.  We also
show that the phase control of ``K${}_{2}$NiF${}_{4}$''-type
La${}_{2}$CuO${}_{4}$ versus ``Nd${}_{2}$CuO${}_{4}$''-type
La${}_{2}$CuO${}_{4}$ can be achieved by varying the synthesis
temperature and using different substrates.
\end{abstract}


\maketitle

\section{INTRODUCTION}

 The rare earth copper oxides of the general chemical formula
RE${}_{2}$CuO${}_{4}$ take two different crystal structures: K${}_{2}$NiF${}_{4}$
(abbreviated as `` T '') and Nd${}_{2}$CuO${}_{4}$ (`` T' '').  The structural
difference between T and T' can be viewed simply as the difference in the RE-O
arrangements: rock-salt-like versus fluorite-like.  With regard to the Cu-O
coordination, however, there is a significant difference: T has octahedral
CuO${}_{6}$, whereas T' has two-dimensional square-planar CuO${}_{4}$. 
Empirically, the former accepts only hole doping, the latter only electron
doping.  The T structure is formed with large La${}^{3+}$ ions, while the T'
structure is formed with smaller RE${}^{3+}$ ions, such as RE = Pr, Nd, Sm, Eu,
and Gd\cite{Nd2CuO4}.  The T-T' boundary lies between
La${}^{3+}$ and Pr${}^{3+}$.  Namely, La${}_{2}$CuO${}_{4}$ is at the borderline of the
T-phase stability.

 The crystal chemistry of the rare earth copper oxides has been explained
in terms of the crystallographic tolerance factor ($t$)\cite{RE2CuO4,(LaLn)2CuO4}, which
is defined as
\begin{equation}
t = \frac{r_{i}({\rm RE^{3+}}) + r_{i}({\rm O^{2-}})}
{\sqrt{2} \times \{r_{i}({\rm Cu^{2+}}) + r_{i}({\rm O^{2-}})\}}
\end{equation}
where {\it r${}_{i}$}(RE${}^{3+}$), {\it r${}_{i}$}(Cu${}^{2+}$), and
{\it r${}_{i}$}(O${}^{2-}$) are the ionic radii for RE${}^{3+}$, Cu${}^{2+}$, and
O${}^{2-}$ ions.  The $t$ values for La${}_{2}$CuO${}_{4}$ and
Pr${}_{2}$CuO${}_{4}$ are evaluated as 0.8685 and 0.8562 using the
room-temperature ionic radii by Shannon and Prewitt\cite{Ionic}.  From
the extensive data collected for a variety of RE${}_{2}$CuO${}_{4}$-type
cuprates, the critical (room-temperature) value for the T
$\rightarrow$ T' transition is presumed to be $t_{\rm c}$ = 0.865, below which T
is unstable\cite{RE2CuO4,(LaLn)2CuO4}.

 The different thermal expansion (``thermal-expansion mismatch'') between the RE-O
and Cu-O bond lengths plays an important role in the T-versus-T' stability as
pointed out initially by Manthiram and Goodenough\cite{Thermal}.  The
``ionic'' RE-O bond has a larger thermal expansion than the ``covalent''
Cu-O bond, which leads to the increase in $t$ with increasing
temperature.  Hence the T phase is stable at high temperatures whereas
the T' phase is stable at low temperatures.  In the case of
La${}_{2}$CuO${}_{4}$, the transition from T to T' is predicted to occur
at around 700 K (427${}^{\circ}$C), where $t$(700 K) $\sim$ 0.88.  There
have been a few attempts to stabilize the T' phase of
La${}_{2}$CuO${}_{4}$ in the past.  However, a conventional solid-state
reaction method requires firing temperature of at least 500${}^{\circ}$C
even with coprecpitated fine powders, so it could not produce
single-phase T'-La${}_{2}$CuO${}_{4}$.  Bulk synthesis of
T'-La${}_{2}$CuO${}_{4}$ has been achieved only by a very special recipe
as given by Chou {\it et al.}\cite{T'}  Their recipe consists of the
following two steps.  The first step is to reduce T-La${}_{2}$CuO${}_{4}$
with hydrogen around 300${}^{\circ}$C and obtain the
Sr${}_{2}$CuO${}_{3}$-like phase.  The second step is to convert the
Sr${}_{2}$CuO${}_{3}$-like phase to T'-La${}_{2}$CuO${}_{4}$ by
reoxygenation below 400${}^{\circ}$C.  The resultant product was the
single-phase T', although x-ray peaks were broadened due to the
considerable lattice disorder and defects.

 In thin-film synthesis, the reaction temperature can be lowered significantly,
since reactants are much smaller in size and also more reactive than in bulk
synthesis.  The reactants in thin-film synthesis are atoms or molecules or ions
or clusters, depending on the technique employed.  The limiting case is achieved
by reactive coevaporation from metal sources, in which the reactants are atoms
and the oxidation reaction is initiated on a substrate.  Using this reactive
coevaporation technique, we have learned from our ten-year experience that
cuprate films crystallize at temperatures as low as 400${}^{\circ}$C.  This enabled us to
synthesize single-phase T'-La${}_{2}$CuO${}_{4}$.  In this article we describe the phase
control of ``K${}_{2}$NiF${}_{4}$''-type La${}_{2}$CuO${}_{4}$ versus
``Nd${}_{2}$CuO${}_{4}$''-type La${}_{2}$CuO${}_{4}$ by varying the synthesis temperature
and using different substrates.

\section{EXPERIMENT}

 We grew La${}_{2}$CuO${}_{4}$ thin films in a customer-designed MBE chamber from metal sources
using multiple electron-gun evaporators with accurate stoichiometry control of
the atomic beam fluxes.  During growth, RF activated atomic oxygen was used for
oxidation.  The chamber pressure during growth was $6 \times 10^{-6}$ Torr.  The
substrate temperature was varied from 425${}^{\circ}$C to 725${}^{\circ}$C.  The growth
rate was $\sim$ 1.5 \AA/s, and the film thickness was typically 450
\AA.  After the evaporation, most of the films were cooled to temperatures lower
than 200${}^{\circ}$C at a rate lower than 20${}^{\circ}$C/min in $1
\times 10^{-5}$ Torr molecular oxygen to avoid phase decomposition.  Some of the films
were cooled in vacuum or in ozone to investigate the change of the transport properties by
excess oxygen.  

 In order to examine the substrate influence on the selective phase
stabilization\cite{(LaNdSr)2CuO4}, we used various substrates as listed in
Table 1.  The in-plane lattice constant ($a_{\rm s}$) covers from 3.6 \AA
\ to 4.2 \AA \ , which should be compared to $a_{0}$ = 3.803
\AA \ for T-La${}_{2}$CuO${}_{4}$ and $a_{0}$ = 4.000 - 4.010 \AA \ for
T'-La${}_{2}$CuO${}_{4}$ (T-La${}_{2}$CuO${}_{4}$ has orthorhombic structure with $a' =
5.3574$ \AA \ and $b' = 5.4005$
\AA, and $a_{0}$ is calculated as $\sqrt{(a' \times b')/2}$).  The crystal
structures include perovskite, K${}_{2}$NiF${}_{4}$, NaCl, and CaF${}_{2}$ (fluorite). 
We deposited films simultaneously on all the substrates listed in Table 1, which were
pasted to one substrate holder by Ag paint.  This avoids run-to-run variations.

 The lattice parameters and crystal structures of the films were determined using
a standard x-ray diffractometer.  Resistivity was measured by the standard
four-probe method using electrodes formed by Ag evaporation.

\section{RESULTS and DISCUSSION}

\subsection{Effect of synthesis temperature on the selective phase stabilization}

 Figure 1 shows the x-ray diffraction (XRD) patterns of La${}_{2}$CuO${}_{4}$ films grown on
NdCaAlO${}_{4}$ (NCAO) substrates with different synthesis temperatures ($T_{\rm s}$). 
Since the $c$-axis lattice constant ($c_{0}$) is distinct between T and T' ($c_{0}$(T) =
13.15 \AA \ versus $c_{0}$(T') = 12.55 \AA), the phase identification is rather
straightforward.  The calculated patterns for T and T' are also included in Fig. 1.  The
films grown at $T_{\rm s}$ $>$ 625${}^{\circ}$C are single-phase T, while the films grown
at $T_{\rm s}$ = 500 - 550${}^{\circ}$C are single-phase T'.  The films grown at $T_{\rm
s}$ = 575 - 600${}^{\circ}$C are a two-phase mixture of T and T' with T' more dominant
for lower $T_{\rm s}$.  The films grown below $T_{\rm s}$ = 475${}^{\circ}$C show
unidentified peaks at 2$\theta$ $\sim$ 31.4${}^{\circ}$ and 65.5${}^{\circ}$.  From this
result, we can see the following trend for synthesis temperature on the
selective phase stabilization.  High $T_{\rm s}$ stabilizes T and low
$T_{\rm s}$ stabilizes T'.

\subsection{Effect of substrates on the selective phase stabilization}

 Figure 2 shows the XRD patterns of La${}_{2}$CuO${}_{4}$ films grown at $T_{\rm s}$ =
525${}^{\circ}$C on different substrates.  Of these films in this figure, the films on
KTaO${}_{3}$ (KTO), NCAO, and ZrO${}_{2}$(Y) (YSZ) are single-phase T'\cite{YSZ}, while
the films on LaSrGaO${}_{4}$ (LSGO), LaAlO${}_{3}$ (LAO), LaSrAlO${}_{4}$ (LSAO),
PrSrAlO${}_{4}$ (PSAO), and NdSrAlO${}_{4}$ (NSAO) are single-phase T.  On YAlO${}_{3}$,
the film is dominantly T' with a trace amount of T.  On SrTiO${}_{3}$ (STO) and
NdGaO${}_{3}$ (NGO), the films are clearly a mixture of T and T'.  The film on STO
contains some amount of the T${}^{\ast}$-like (!) phase\cite{T*}.  On MgO (MGO), no clear
peak is observed.  The
$c_{0}$ values of these films together with films on other substrates are summarized in
Fig. 3.  Because of epitaxial strain\cite{strain}, $c_{0}$ of the T
structure is noticeably substrate-dependent: the longest ($c_{0}$ = 13.25
\AA) for LSAO and the shortest ($c_{0}$ = 13.05 \AA) for LSGO.  From
these results, we can see the following trend for a substrate lattice
parameter on the selective phase stabilization.  Substrates with
$a_{0}$ of 3.70 - 3.85 \AA \ stabilize T, and substrates with $a_{0}$ of $>$ 3.90
\AA \ or $<$ 3.70 \AA \ stabilize T' (or destabilize T).

 Next we mention the effect of substrate crystal structure on the selective phase
stabilization.  If, in Fig. 2, one compares the films grown on perovskite and
K${}_{2}$NiF${}_{4}$-type substrates with almost the same $a_{0}$, for example, NGO
($a_{0}$ = 3.838 \AA) vs LSGO ($a_{0}$ = 3.843 \AA) or YAO ($a_{0}$ = 3.715 \AA) vs NSAO
($a_{0}$ = 3.712 \AA), one can notice the trend that
K${}_{2}$NiF${}_{4}$-type substrates have a tendency to stabilize the T
structure rather than the T' structure.

\subsection{Phase diagram in the $T_{\rm s}$-$a_{\rm s}$ plane}

 Our survey was performed at $T_{\rm s}$ from 425${}^{\circ}$C to 725${}^{\circ}$C on
all substrates in Table 1.  Figure 4 summarizes the results, which show the
phase diagram on the selective stabilization of T versus T' in the
$T_{\rm s}$-$a_{\rm s}$ plane.\\
{\it High $T_{\rm s}$ (625 $\sim$ 725${}^{\circ}$C)}\\
The films on most of the substrates are single-phase T.  There are three
exceptional substrates: KTO, YAO, and YSZ.  The films on KTO and YSZ do
not show any definite x-ray peak. The film on YAO is a mixture of T and
T' even at the highest temperature investigated.  This can be explained by
interdiffusion of Y from YAO substrates into La${}_{2}$CuO${}_{4}$ since
Y substitution for La is known to stabilize the T' structure.\\
{\it Low $T_{\rm s}$ (450 $\sim$ 600${}^{\circ}$C)}\\
The films on the T-lattice matched substrates (LSGO, LAO, LSAO, PSAO, and
NSAO) are single-phase T.  The films on T'-lattice matched KTO and on
fluorite YSZ are single-phase T'.  The films on other substrates (STO,
NGO, YAO, and NCAO) are a two-phase mixture of T and T' with T' more
dominant for lower
$T_{\rm s}$.

\subsection{Comparison of T-La${}_{2}$CuO${}_{4}$ and T'-La${}_{2}$CuO${}_{4}$}

 Next, we make a brief comparison of the physical properties of T-La${}_{2}$CuO${}_{4}$ and
T'-La${}_{2}$CuO${}_{4}$, which have the same chemical formula but different crystal
structures.  Figure 5 shows the temperature dependences of resistivity for both
the phases.  The solid lines represent the $\rho$ - $T$ curves for the
films cooled in vacuum to ambient temperature, which do not have excess
oxygen but might have slight oxygen deficiencies
(La${}_{2}$CuO${}_{4+\delta}$ with $\delta \sim 0$).  The broken lines
represent those for the films cooled in ozone, which have interstitial
excess oxygen ($\delta > 0$)\cite{ozone}.  The excess oxygen occupies the
tetrahedral site in T, and the apical site in T'.  The vacuum-cooled T
film has much higher resistivity (by several orders of magnitudes at low
temperatures) than the vacuum-cooled T' film.  In fact,
T'-La${}_{2}$CuO${}_{4}$ is metallic down to 180 K\cite{metallicT'}.  The ozone
cooling causes a totally opposite effect on T and T'.  The resistivity of the T film gets
lowered by five orders of magnitudes at room temperature from $\sim 50 \ 
\Omega \cdot {\rm cm}$ to $\sim 5 \times 10^{-4} \ \Omega \cdot {\rm cm}$, indicating
that holes doped by excess oxygen are itinerant.  Furthermore the film becomes
superconducting.  In contrast, the resistivity of the T' film increases, indicating
that holes doped by excess oxygen are localized\cite{metallicT'}.

\section{SUMMARY}

 In summary, we have demonstrated that La${}_{2}$CuO${}_{4}$ can crystallize in the Nd${}_{2}$CuO${}_{4}$
structure using low-temperature thin-film synthesis.  Furthermore the phase
control of ``K${}_{2}$NiF${}_{4}$''-type La${}_{2}$CuO${}_{4}$ versus
``Nd${}_{2}$CuO${}_{4}$''-type La${}_{2}$CuO${}_{4}$ can be achieved by varying the
synthesis temperature and also the substrate.   The general trends are as follows: (i)
high $T_{\rm s}$ stabilizes T and low $T_{\rm s}$ stabilizes T', (ii) substrates with
$a_{\rm s}$ $\sim$ 3.70 - 3.85
\AA \ stabilize T and substrates with $a_{\rm s}$ $>$ 3.90 \AA \ or $a_{\rm s}$ $<$ 3.70
\AA \ stabilize T' (or destabilize T), and (iii) K${}_{2}$NiF${}_{4}$-type substrates
stabilize T.

\begin{acknowledgments}
The authors thank Dr. H. Sato, Dr. H. Yamamoto, Dr. S. Karimoto, and Dr. T.
Yamada for helpful discussions, and, Dr. H. Takayanagi and Dr. S. Ishihara for
their support and encouragement throughout the course of this study.
\end{acknowledgments}


\begin{thebibliography}{12}
\bibitem{YSZ} On YSZ, judging from the peak positions, the film seems to be single-phase
T', while the peak intensity ratios do not agree with the calculated ratios.
\end{thebibliography}


\begin{figure}
\includegraphics{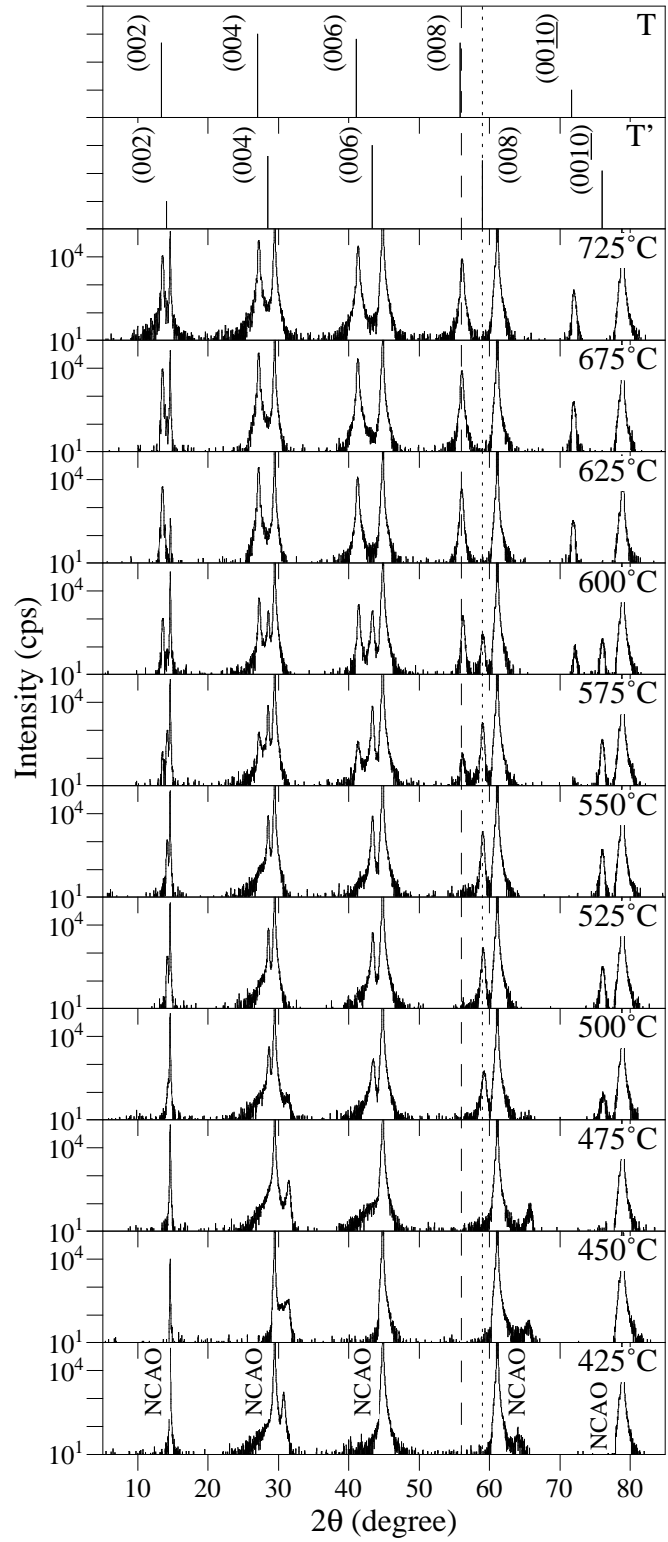}
\caption{XRD patterns for La${}_{2}$CuO${}_{4}$ films grown on NCAO
substrates at $T_{\rm s}$ = 725 - 425${}^{\circ}$C.  The top two patterns
are simulations for the T and T' structure.  The broken and dotted lines
indicate the peak positions of the (008) line for the T and T'
structure, respectively.  Peak positions of NCAO are indicated in the lowest
figure.}
\end{figure}

\begin{figure}
\includegraphics{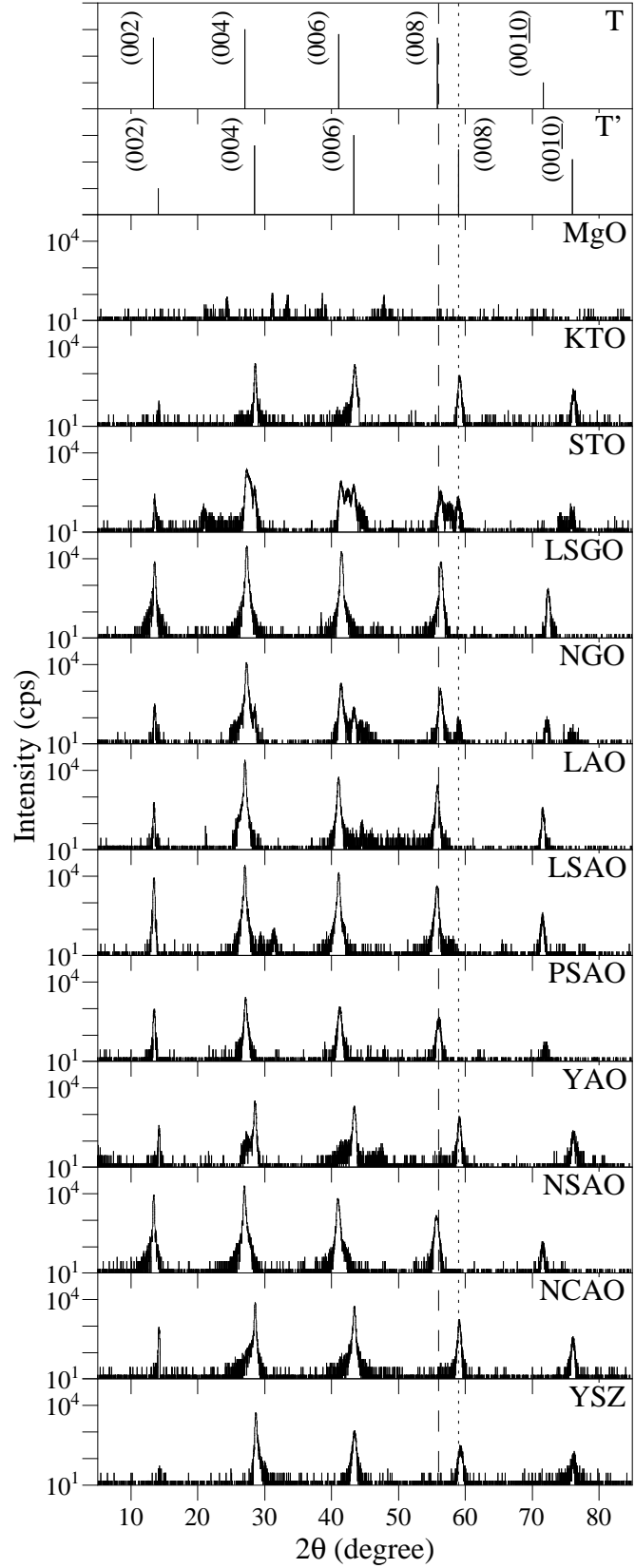}
\caption{XRD patterns for La${}_{2}$CuO${}_{4}$ films grown on various
substrates at $T_{\rm s}$ = 525${}^{\circ}$C.  The top two patterns are
simulations for the T and T' structure.  Substrate peaks are
removed.  The broken and dotted lines indicate the peak positions of the
(008) line for the T and T' structure, respectively.}
\end{figure}

\begin{figure}
\includegraphics{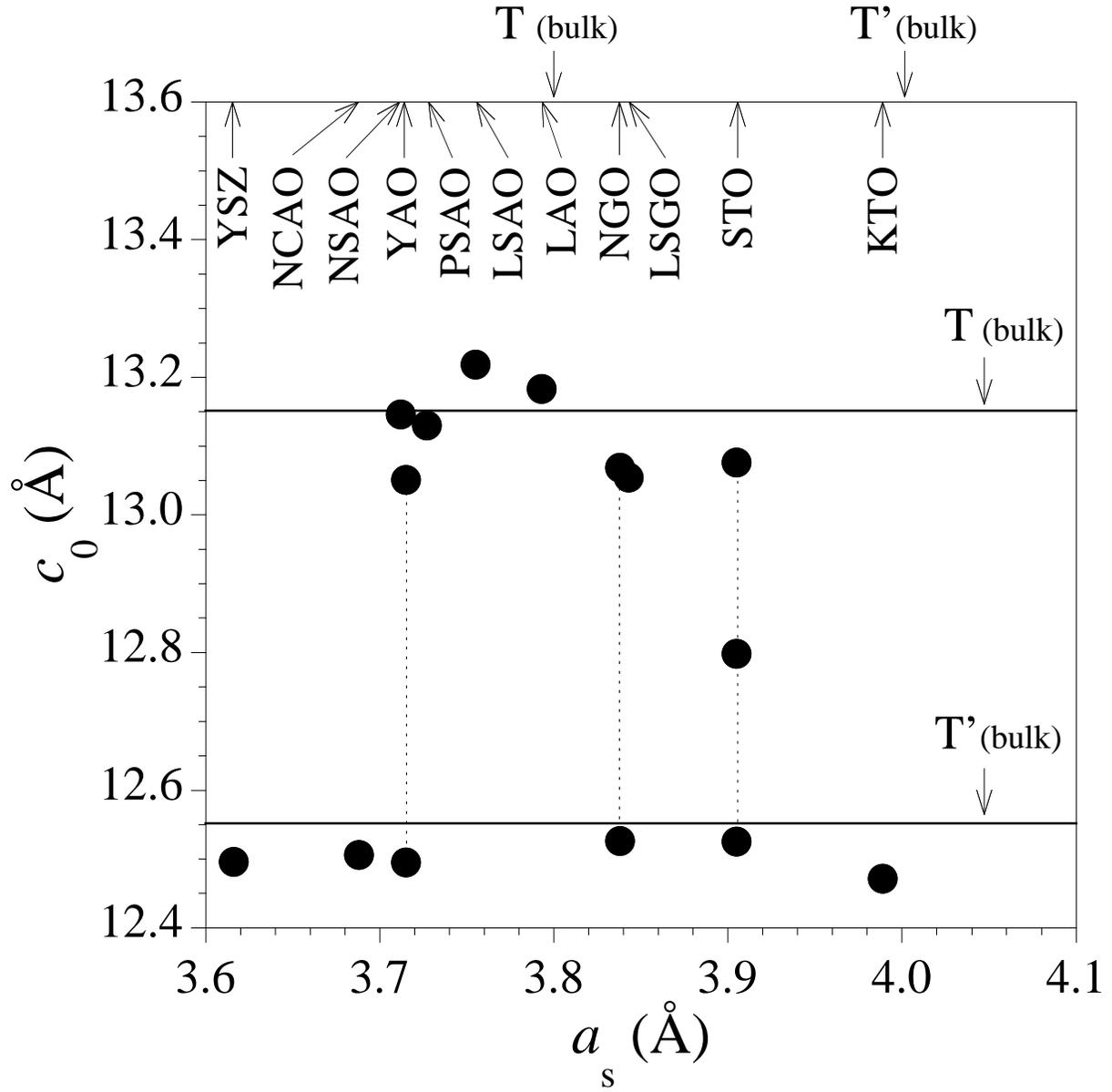}
\caption{Film's $c_{0}$ versus substrateÕs $a_{\rm s}$ for
La${}_{2}$CuO${}_{4}$ films grown at $T_{\rm s}$ = 525${}^{\circ}$C on
different substrates.  The lattice constants of bulk
T- and T'-La${}_{2}$CuO${}_{4}$ ($a_{0}$ = 3.803 \AA, $c_{0}$ = 13.15
\AA \ for T and $a_{0}$ = 4.005 \AA, $c_{0}$ =12.55 \AA \ for T') are
indicated by arrows together with $a_{\rm s}$ of the substrates.  The
circles connected by the vertical dotted lines indicate multi-phase
formation.  The $c_{0}$ values of the T structure is noticeably
substrate-dependent because of epitaxial strain: the longest ($c_{0}$ =
13.25 \AA) for LSAO and the shortest ($c_{0}$ = 13.05 \AA) for LSGO. 
The $c_{0}$ value of 12.8 \AA \ on STO seems to correspond to the
T${}^{*}$-like phase.}
\end{figure}

\begin{figure}
\includegraphics{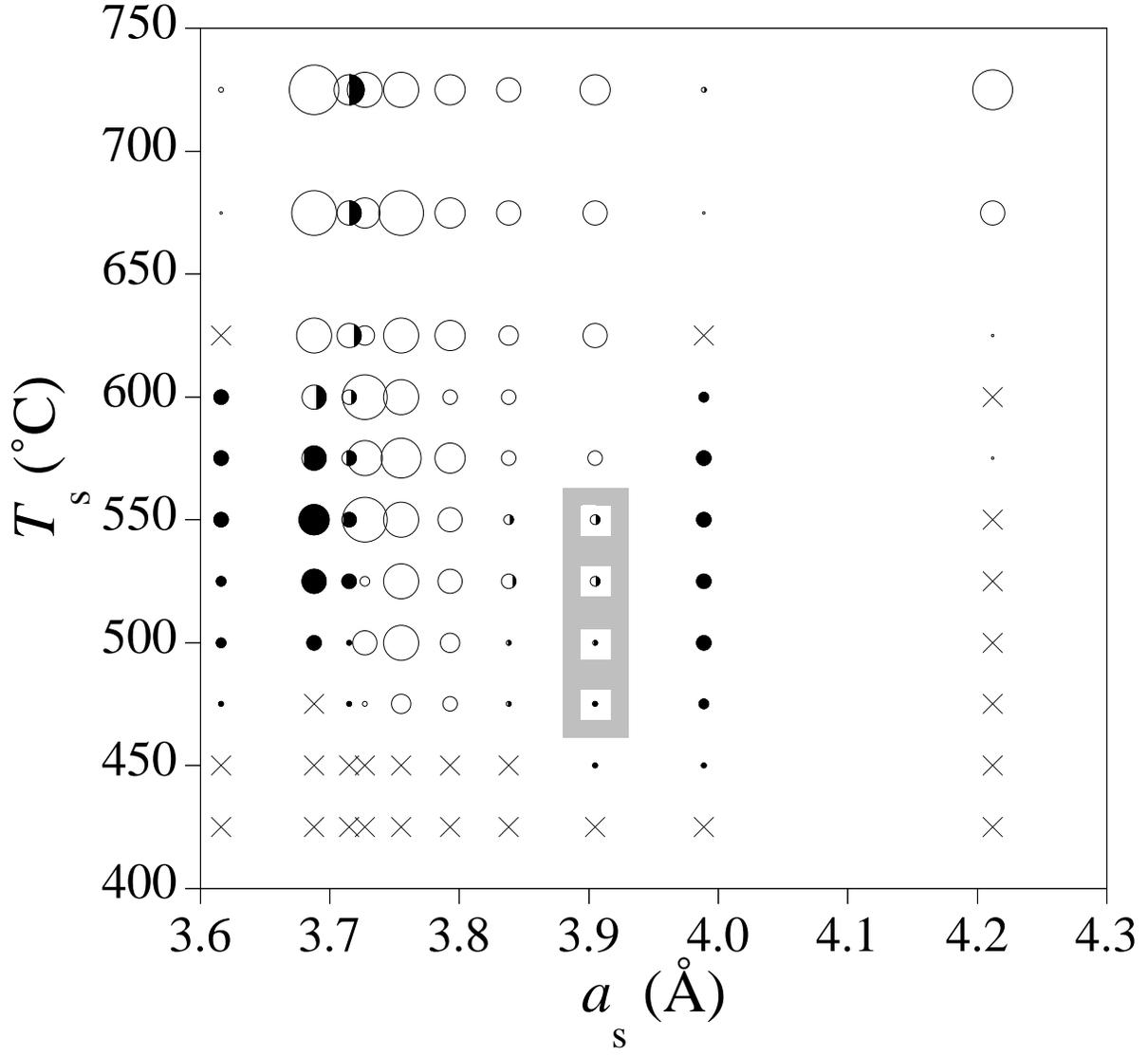}
\caption{Phase diagram on the selective stabilization of T versus T' in
the $T_{\rm s}$ - $a_{\rm s}$ plane.  The crosses indicate no phase
formation.  The open circles represent single-phase T while the filled
circles represent single-phase T'.  The partially filled circles
represent a two-phase mixture.  The size (area) of the circles is
proportional to the XRD peak intensities of the (006) lines.  For
two-phase mixed films, the ratio of the unshaded and the shaded areas
represent the ratio of the T and T' peak intensity of the (006) line. 
The results on LSGO and NSAO substrates are not included to avoid
overlapping with the results on NGO and YAO.  On LSGO and NSAO, the T
structure is formed for 725${}^{\circ}$C $>$ $T_{\rm s}$ $>$
475${}^{\circ}$C, and the T' structure is not formed for any $T_{\rm
s}$.  The gray area at $a_{\rm s}$ = 3.905 \AA \ (STO) indicate the formation of the
T${}^{*}$-like phase.}
\end{figure}

\begin{figure}
\includegraphics{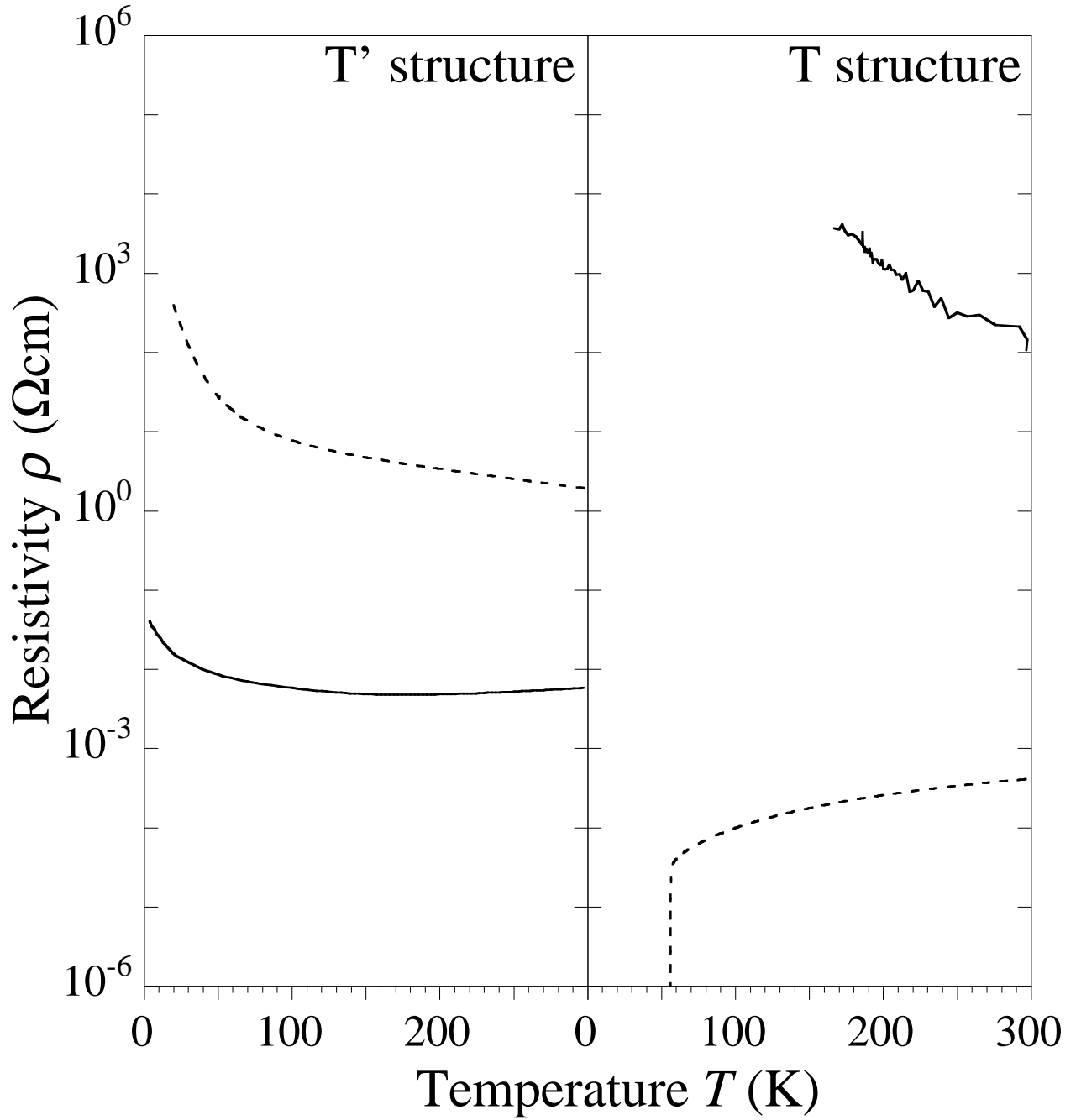}
\caption{Comparison of resistivity ($\rho$) - temperature ($T$)
curves between T-La${}_{2}$CuO${}_{4+\delta}$ and
T'-La${}_{2}$CuO${}_{4+\delta}$ films.  The solid lines are for films
cooled in vacuum ($\delta \sim 0$) while the broken lines are for films
cooled in ozone ($\delta > 0$).  With vacuum cooling, the T film has much
higher resistivity than the T' film.  Ozone cooling causes a totally
opposite effect on T and T': the T film gets metallic and superconducting
whereas the T' film gets more insulating.}
\end{figure}

\begin{table}
\caption{Crystal structure and $a$-axis lattice constant ($a_{\rm s}$)
for the substrates used in this work.  The in-plane lattice constants ($a_{0}$) for
T'-La${}_{2}$CuO${}_{4}$ and T-La${}_{2}$CuO${}_{4}$ are also included.}
\begin{ruledtabular}
\begin{tabular}{lccc}
\ Substrate & Abbreviation & $a_{\rm s}$ or $a_{0}$ (\AA) & Crystal structure\\
\hline
\ MgO (100)                & MGO  & 4.212 & NaCl \\
\ KTaO${}_{3}$ (100)       & KTO  & 3.989 & perovskite \\
\ SrTiO${}_{3}$ (100)      & STO  & 3.905 & perovskite \\
\ LaSrGaO${}_{4}$ (001)    & LSGO & 3.843 & K${}_{2}$NiF${}_{4}$ \\
\ NdGaO${}_{3}$ (100)      & NGO  & 3.838 & perovskite \\
\ LaAlO${}_{3}$ (100)      & LAO  & 3.793 & perovskite \\
\ LaSrAlO${}_{4}$ (001)    & LSAO & 3.755 & K${}_{2}$NiF${}_{4}$ \\
\ PrSrAlO${}_{4}$ (001)    & PSAO & 3.727 & K${}_{2}$NiF${}_{4}$ \\
\ YAlO${}_{3}$ (100)       & YAO  & 3.715 & perovskite \\
\ NdSrAlO${}_{4}$ (001)    & NSAO & 3.712 & K${}_{2}$NiF${}_{4}$ \\
\ NdCaAlO${}_{4}$ (001)    & NCAO & 3.688 & K${}_{2}$NiF${}_{4}$ \\
\ ZrO${}_{2}$ (Y) (100)    & YSZ  & 3.616 & fluorite \\
\hline
\ T'-La${}_{2}$CuO${}_{4}$ &      & 4.005 & Nd${}_{2}$CuO${}_{4}$ \\
\ T-La${}_{2}$CuO${}_{4}$  &      & 3.803 & K${}_{2}$NiF${}_{4}$ \\
\end{tabular}
\end{ruledtabular}
\end{table}

\end{document}